\documentclass{llncs}
\usepackage{graphicx}
\usepackage{eurosym}
\usepackage[misc]{ifsym}
\usepackage{verbatim}
\usepackage{appendix}
\usepackage{url}
\begin{document}

\title{A Framework for Automatic Monitoring of Norms that regulate Time Constrained Actions\thanks{Funded by the SNSF (Swiss National Science Foundation) grant no. 200021\_175759/1.} \fnmsep \thanks{Proc. of the COINE 2021 co-located with AAMAS 2021, 3rd May 2021, London, UK. All Rights Reserved.}
}

\author{Nicoletta Fornara(\Letter)\inst{1}, Soheil Roshankish \inst{1}, Marco Colombetti\inst{2}}
\institute{Universit\`{a} della Svizzera italiana, via G. Buffi 13, 6900 Lugano, Switzerland\\
\email{nicoletta.fornara@usi.ch, soheil.roshankish@usi.ch},  \and
Politecnico di Milano, piazza Leonardo Da Vinci 32, Milano, Italy \\
\email{marco.colombetti@polimi.it}}

\maketitle
\begin{abstract}
This paper addresses the problem of proposing a model of norms and a framework for automatically computing their violation or fulfilment. The proposed T-NORM model can be used to express abstract norms able to regulate classes of actions that should or should not be performed in a temporal interval. We show how the model can be used to formalize obligations and prohibitions and for inhibiting them by introducing permissions and exemptions. The basic building blocks for norm specification consists of rules with suitably nested components. The activation condition, the regulated actions, and the temporal constrains of norms are specified using the W3C Web Ontology Language (OWL 2). Thanks to this choice, it is possible to use OWL reasoning for computing the effects that the logical implication between actions has on norms fulfilment or violation. The operational semantics of the T-NORM model is specified by providing an unambiguous procedure for translating every norm and every exception into production rules.
\end{abstract}

\section{Introduction}\label{sec:introduction}
In this paper, we present the T-NORM model, a model for the formalization of \textit{relational norms} that regulate classes of actions that agents perform in the society and that put them in relation to other agents, like for example paying, lending, entering in a limited traffic area, and so on. Our proposal is strictly related to the specification of the operational semantics of such a model, this to make it possible to provide monitoring and simulation services on norms specifications. Specifically, the proposed model can be used for automatically computing the fulfilment or violation of active (or in force) obligations and prohibitions formalized for regulating a set of actions that should or should not be performed in a \textit{temporal interval}. The fact that regulated actions are time constrained is an important characteristic of the proposed model. Another important aspect is that once a set of obligations and prohibitions are formalized they may be further refined by the definition of permissions and exemptions.

In NorMAS literature \cite{NORMAS:2018} it is possible to find numerous formal models for the specification of norms, contracts, commitments and policies. Many of them can be used for regulating the performance of actions or for requiring the maintenance of certain conditions, but very often those actions and conditions may only be expressed using propositional formulae. This choice makes it difficult to express the relation between the regulated actions (or conditions) and time. This is an important limit in the expressiveness of those models, because there are numerous real examples of norms and policies whose relation with time intervals is important for their semantics; for example in e-commerce, deadlines (before which a payment must be done) are fundamental for computing the fulfilment or violation of contracts.

When temporal aspects are important for the specification of norms and for reasoning on their evolution in time, one approach may consist in using temporal logics (for example Linear Temporal Logic LTL) for expressing and reasoning on time related constrains. But this choice presents important limits when automatic reasoning is required for computing the time evolution of the normative state, as discussed in \cite{Panagiotidi:2013:TNA:2731087.2731112}.

In our approach we propose to formalize some components of the norms, i.e. their activation condition, the regulated actions, and the temporal constrains using semantic web languages, specifically the W3C Web Ontology Language (OWL 2)\footnote{\url{https://www.w3.org/TR/owl2-overview/}}. This for two reasons. First, it should be easier for those who want to formalize norms with our model to use a standard language that is fairly well known and taught in computer science graduate courses. Second, this language has a formal semantics on which automatic reasoning can be performed. Moreover, OWL is more expressive than propositional formulae\footnote{Description Logics (DLs), which are a family of class (concept) based knowledge representation formalisms, are more expressive than propositional logic, and they are the basis for ontology languages such as OWL \cite{10.1007/11564751_2Horrocks:OWL}.}, which are used in many other norm models. The idea of formalizing policies using semantic web languages is spreading also thanks to the success of the ODRL (Open Digital Rights Language) policy expression language\footnote{\url{https://www.w3.org/TR/odrl-model/}}, which has been a W3C Recommendation since February 2018.

Our idea is to propose a model of norms that norm designers can use for formalizing the intuitive meaning of having an obligation or a prohibition. That is, when something happens and certain conditions hold, an agent is obliged or prohibited to do something in a given interval of time. What is innovative with respect to other approaches is that instead of explicitly labelling a norm as an obligation or a prohibition, we let the norm designer explicitly express what sequence of events will bring to a violation or a fulfilment. To do so the norm designer needs to be able to describe triggering events or actions and their consequences. The resulting model will let the norm designer to specify norms as \textit{rules} nested into each other.

The main contributions of this paper are: (i) the definition of a model of norms that can be used to specify numerous types of deontic relations, i.e. different types of obligations and prohibitions, and their exceptions; (ii) the definition of its operational semantics (by combining OWL reasoning and forward chaining) that can be used to automatically monitor or simulate the fulfillment or violation of a set of norms; (iii) the proposal of a set of different types of concrete norms, that can be used to evaluate the expressive power of a norms model.

This paper is organized as follows: in Section \ref{sec:goal} the main goals that guided the design of the norms model are presented. In Section \ref{sec:model:norms} the T-NORM model of norms is introduced and in Section \ref{sec:semantics} it operational semantics is provided. Finally, in Section \ref{sec:related:work} the proposed model is compared with other existing approaches.

\section{Design Goals}\label{sec:goal}
In this section we list the main goals that guided us in the design of the proposed model and in the definition of its operational semantics.

Our first goal is to propose a model of norms able to regulate \textit{classes of actions}; for example we want to be able to formalize a norm that regulates all the accesses to a limited traffic zone of a metropolis and not only the access of a specific agent. This objective is not satisfied by those models, like ODRL and all its extensions or profiles \cite{DBLP:journals/aicom/FornaraC19,2019:ODRL:DeVosKirranePadget} where the policies designer has to specify the specific debtor of every policy instance.

Our second goal is to define a model of norms able to regulate classes of actions whose performance is \textit{temporally constrained}. For instance, the following norm regulates all the access to an area and the subsequent action of paying is temporally constrained by a deadline, which in turn depends on the access instant of time: \textit{``when an agent enters in the limited-traffic area of Milan, between 7 a.m. and 7 p.m., they have to pay 6 euro within 24 hours''}. The first and the second goal will bring us to define a model for expressing norms that may be applied multiple times to different agents and may be activated by numerous events happening in different instant of time.

Starting from the experience that we gained by using the model of obligations, prohibitions and permits presented in our previous paper \cite{DBLP:journals/aicom/FornaraC19}, we have developed a third goal for our model. The goal is to propose a model made of basic constructs that can be combined by the norm designer to express different types of deontic relations without the need to introduce a pre-defined list of deontic types, like obligation, prohibition, and permission. This has the advantage that whenever a new kind of norms is required, like for example the notion of exemption or the notion of right, there is no need to introduce into the model a new type with its operational semantics. With the model proposed in this paper, it is possible to use few basic constructs and combine them in different ways to express the obligation to perform an action before a given deadline or the prohibition to perform an action within an interval of time. Our idea is to allow the norm designers to explicitly state what behaviour will bring to a violations and what behaviour will bring to a fulfilment, regardless of they are formalizing an obligation or a prohibition. Moreover, in this new model, permissions are not treated any more as first-class objects, but they are formalized as exceptions to prohibitions, while exemptions are formalized as exceptions to obligations.

Our fourth goal is to provide an operational semantics of our model of norms that will make it possible to \textit{monitor} or \textit{simulate} the evolution of their state in time. Our goal is mainly to be able to automatically compute if a policy is \textit{active} (or in-force) and then if it becomes \textit{fulfilled} or \textit{violated} on the basis of the events and actions performed by agents. Monitoring is crucial from the point of view of policy's \textit{debtor} for checking if their behaviour is compliant and it is relevant for policy's \textit{creditors} to react to violations. Simulation may be used for evaluating in advance the effects of the performance of certain actions. Another useful service that can be provided on a set of policies is checking their consistency. Checking for example if a given action is contemporarily obliged and prohibited. This can be done at design time by using model-checking techniques, but it is not among the objectives of our model. Nevertheless, by proposing a model of norms that can be tracked in its evolution in time it will be possible to detect inconsistencies at run-time in every instant of time of the monitoring or simulation process.

\section{Model of norms}\label{sec:model:norms}
The idea that has guided us in the definition of the T-NORM model is to give norm designers a tool to describe which sequence of events or actions would bring an agent to the violation or to the fulfilment of a norm. This approach has the advantage of providing norm designers with a model that in principle can be used to define any type of deontic relationship, like obligations, prohibitions, permissions, exemptions, rights and so on. This is a crucial difference with respect to the models having a pre-defined set of deontic types, like it is the case for ODRL, OWL-POLAR \cite{DBLP:journals/ws/SensoyNVS12}, and also our previous proposal of a model for monitoring obligations, prohibitions, and permissions \cite{DBLP:journals/aicom/FornaraC19}.

The intuitive meaning of having an obligation (resp. a prohibition) that we want to capture is the following one: when an activation event happens and some contextual conditions are satisfied, it is necessary to compute some parameters (for example the deadline) and to start to monitor the performance of a specific regulated action or class of actions. In turn, if an action, which matches the description of the regulated one, is performed before another event (for example a time event that represents a deadline), then the obligation is fulfilled (resp. the prohibition is violated); otherwise, if the regulated action cannot be performed anymore (for example because the deadline has elapsed) the obligation is violated (resp. the prohibition is fulfilled).

To capture this intuitive meaning we decided to represent norms as rules that determine the conditions under which a fulfilment or violation is generated. In what follows we show how different types of norms can be represented in this way. Then in the next section we show how our norms can be translated into production rules, thus assigning an unambiguous operational semantics to our norm formalism.

The idea of representing norms in the form of rules is not new in NorMAS literature \cite{Garcia-Camino:2005:INE:1082473.1082575}. However, in order to explicitly specify the sequence of events that bring to a violation or a fulfilment, we propose a model of norms where the basic building blocks for norm specification consists of rules with suitably nested components. Thanks to this choice, as we will discuss in Section \ref{sec:semantics}, the operational semantics of our model of norms can be easily expressed using productions. Our idea is to express the meaning of having an obligation or a prohibition by nested rules of the following form:

\footnotesize
\begin{verbatim}
NORM Norm_n
[ON ?event1 WHERE conditions on ?event1
THEN
   COMPUTE]
   CREATE DeonticRelation(?dr);
   ASSERT isGenerated(?dr,Norm_n); [activated(?dr,?event1);]
   ON ?event2 [BEFORE ?event3 WHERE conditions on ?event3]
      WHERE actor(?event2,?agent) AND conditions on ?event2
   THEN ASSERT fulfills(?agent,?dr); fullfilled(?dr,?event2)|
               violates(?agent,?dr); violated(?dr,?event2)
  [ELSE ASSERT violates(?agent,?dr); violated(?dr,?event3)|
               fulfills(?agent,?dr); fulfilled(?dr,?event3]
\end{verbatim}
\normalsize

In the proposed model the first (optional) \texttt{ON...THEN} component is used for expressing \textit{conditional norms}, i.e. norms that start to be in force when a certain event happens and where the \textit{temporal relation} between the activating event and the regulated action is crucial in the semantics of the norm. For example in \texttt{Norm01} (\textit{``when an agent enters in the limited-traffic area of Milan between 7 a.m. and 7 p.m., they have to pay 6 euro within 24 hours''}) the event of entering in the limited-traffic area must occur for the obligation to pay to become active; moreover the entering instant is fundamental for computing the deadline of the payment. The second \texttt{ON...THEN} component is used for expressing the actions regulated by the norm and the consequences of their performance or non-performance.

In the T-NORM model, a norm activation can generate many different \textit{deontic relations}. In other approaches, like for example in \cite{DBLP:conf/atal/Alvarez-NapagaoAVD10}, a norm generates norm instances. We prefer to use the term deontic relation because it can also be used to denote obligations and prohibitions that are not created by activating a norm, but for example by making a promise or accepting an agreement.

In the \texttt{ON ?event WHERE} and in the \texttt{BEFORE ?event WHERE} components, the norm designer has to describe the \textit{conditions} that a real event has to satisfy to match a norm. In our model all the relevant events are represented by data specified in the \textit{State Knowledge Base}. Data for managing the evolution of the state of norms, for example the deontic relation objects, are stored in the \textit{Deontic Knowledge Base}. Obviously, the formalism chosen for representing the data in the \textit{State KB} and the \textit{Deontic KB} determines the syntax for expressing: \textit{norm conditions}, which are evaluated on the \textit{State KB}, and norm actions, which are performed on the \textit{Deontic KB}. Differently from other approaches, where the context or state of the interaction is represented by using propositional formulae \cite{DBLP:conf/atal/Alvarez-NapagaoAVD10,Governatori:2010:10.5555/1862330.1862332}, we decided to formalize the \textit{State KB} and the \textit{Deontic KB} by using semantic web technologies, in particular the W3C Web Ontology Language (OWL 2). This choice has the following advantages:
 \begin{itemize}
   \item events and actions are expressed with a more expressive language, indeed OWL 2 is a practical realization of a Description Logic known as \textit{SROIQ(D)}, which is more expressive than propositional logic;
   \item in the definition of the conceptual model of the State KB it is possible to reuse existing OWL 2 ontologies making the various systems involved in norm monitoring interoperable;
   \item it is possible to perform automatic reasoning (OWL 2 can be regarded as a decidable fragment of First-Order Logic (FOL)) on the \textit{State KB} and deducing knowledge from the asserted one. For example, it is important when the execution of an action logically implies another one. For example, the reproduction of an audio file implies its use, therefore if the use is forbidden so is its reproduction. This is a crucial advantage because instead of creating special properties to relate the regulated actions (like for example it has been done in ODRL with the \texttt{implies}  property\footnote{\url{https://www.w3.org/TR/odrl-model/\#action}}) it is enough to reason on the executed action by using OWL 2 reasoners.
  \end{itemize}

In the specification of \textit{norm conditions}, OWL classes are represented using unary predicates starting with a capital letter and OWL properties are represented using binary predicates starting with a lowercase letter. If an event is described with more conditions, they are evaluated conjunctively, variables (starting with ?) are bound to a value, and a negated condition is satisfied if there is no element in the KB that matches it. In the example reported in this paper, the conceptual model of the events represented in the \textit{State KB} is formalized with the \textit{Event Ontology}in OWL \cite{fornara:SASFA:2011,DBLP:journals/aicom/FornaraC19}, which imports the \textit{Time Ontology} in OWL\footnote{\url{https://www.w3.org/TR/owl-time/}} used for connecting events to instants or intervals of time, and the \textit{Action Ontology} for representing domain-specific actions, like the \textsf{PayAction} class. The conceptual model of the \textit{Deontic KB} is formalized with the \textit{T-Norm Ontology} in OWL\footnote{\url{https://people.lu.usi.ch/fornaran/2021-T-NORM/OWLEventNormOntology.html}}.

In the second component of norms, the \texttt{BEFORE} condition and the \texttt{ELSE} branch are optional. The \texttt{BEFORE} part is mainly used for expressing deadlines for obligations. Although an obligation without a deadline cannot be violated and therefore it is not an incentive to perform the obligatory action, the \texttt{BEFORE} part is not compulsory. The \texttt{ELSE} branch is followed when the regulated action cannot happen anymore in the future, for example if it has to happen before a given deadline (in this case event3 is a time event) and the deadline expires without event2 being performed. In principle other conditions, beside \texttt{BEFORE}, can be used to express other temporal operators but they are not introduced in this version of our model. In the consequent (\texttt{THEN}) parts of a norm, it is possible to specify that:
\begin{itemize}
  \item (\texttt{COMPUTE}) the value of some variables (for example the deadline that depends on the activation time) are computed using arithmetic operations and the value of variables obtained when matching the antecedent;
  \item (\texttt{CREATE}) new individuals belonging to a certain class and having certain properties have to be created in the \textit{Deontic KB} for making the monitoring of norms feasible. Every conditional norm when is activated generates many different deontic relations;
  \item (\texttt{ASSERT}) the value of certain properties of existing individuals created by the norm may be set.
\end{itemize}

The \textit{debtor} of a deontic relation is the agent that is responsible for its violation or fulfilment; usually it is the actor of the regulated action. In legal systems there are exceptions to this general rule (for example for actions performed by minor or people with mental impairment or in cases of strict liability), but we leave this aspect for future works. Specifying the debtor is important because it is the agent to whom sanctions will apply.

The \textit{creditor} of a deontic relation is the agent to whom the debtor owns the performance or non-performance of the regulated action. In certain cases it may be difficult to establish the creditor of a deontic relation (for example who is the creditor of the prohibition of running a red light?); we leave for future works the analysis of this aspect.

\subsection{Expressive power of the model}\label{sec:expressive:power}

By using the T-NORM model we are able to express different types of norms. First of all it is possible to formalize conditional and direct (or un-conditional) obligations and conditional and direct prohibitions. Moreover, every conditional norm (whether it is an obligation or a prohibition) when activated will bring to the creation of \textit{specific deontic relations} or to the creation of \textit{general deontic relations}. Every specific deontic relation regulates the performance of an action by a specific agent, differently every general deontic relation regulates the performance of a class of actions that can be concretely realized by different agents, and therefore can generate many violations or fulfilments.

There exist models, which are not focused on regulating time-constrained actions, where (coherently with deontic logics) prohibitions are merely formalized as obligations to not perform the regulated action. However, when the regulated actions are time constrained it is crucial to react to their performance but also to their non-performance in due time. Think for example to the prohibition expressed by \texttt{Norm02}: \textit{``Italian libraries cannot lend DVDs until 2 years are passed from the distribution of the DVD''}. This prohibition cannot be expressed as an obligation to not lend certain DVDs in a specific time interval, because while an obligation is fulfilled by the performance of a single instance of an action, a prohibition is not fulfilled by a single instance of refraining from the performance of an action. In the T-NORM model the main difference between obligations and prohibitions is that the performance of the regulated action brings about a fulfilment in the case of obligations and a violation in the case of prohibitions.

To illustrate the flexibility of our model, we shall now present examples of different types of norms. Due to space limitation we will not formalize them all. \texttt{Norm01} is an example of \textit{conditional obligation} and each one of its activations creates one \textit{specific} deontic relation to pay 6 euro for the owner of every vehicle that entered in the limited-traffic area. \texttt{Norm01} may be formalized with the T-NORM model in following way:

\footnotesize
\begin{verbatim}
NORM Norm01
ON ?e1
   WHERE RestrictedTrafficAreaAccess(?e1) AND vehicle(?e1,?v) AND
   owner(?v,?agent) AND atTime(?e1,?inst1) AND
   inXSDDateTimeStamp(?inst1,?t1) AND ?t1.hour>7 a.m. AND ?t1.hour<7p.m
THEN
   COMPUTE ?tend_n.hour=?t1.hour+24
   CREATE  DeonticRelation(?dr);TimeEvent(?tevend_n);Instant(?instend_n);
   ASSERT  isGenerated(?dr,Norm01);activated(?dr,?e1);debtor(?dr,?agent);
           end(?dr,?tevend_n); atTime(?tevend_n,?instend_n);
           inXSDDateTimeStamp(?instend_n,?tend_n);
   ON  ?e2  BEFORE ?tevend_n
      WHERE PayAction(?e2) AND reason(?e2,?e1) AND recipient(?e2,'Milan')
      AND price(?e2,6) AND priceCurrency(?e2,euro) AND actor(?e2,?agent1)
   THEN	ASSERT fulfills(?agent,?dr); fulfilled(?dr,?e2)
   ELSE	ASSERT violates(?agent,?dr); violated(?dr,?tevend_n)
\end{verbatim}
\normalsize


\texttt{Norm02} is an example of \textit{conditional prohibition}, its activation creates a \textit{general} deontic relation every time a new DVD is distributed. The general deontic relation created for each specific DVD regulates the actions of all the agents registered in Italian libraries.

The third type of norms is a \textit{conditional prohibition} that generates \textit{specific} deontic relations, an example of this type of norm is \texttt{Norm03}: \textit{``a person who has a positive swab to Covid-19 cannot leave the house for the next 15 days''}. The fourth type of norm is a \textit{conditional obligation} that generates \textit{general} deontic relations, like for example in \texttt{Norm04}: \textit{``when the school bell rings, students have 5 minutes to enter their classroom''}.

An example of \textit{unconditional prohibition} is given by \texttt{Norm05}: \textit{``when the red light is on it is prohibited to pass the traffic light''}. This prohibition is unconditional because there is no need to react to its activation by performing specific actions. For enforcing this prohibition it is enough to check the state of the red light every time an agent pass the traffic light and if the red light is on there is directly a violation. Finally, the following \texttt{Norm06}: \textit{``the lecturer of a course has to organize 2 exams per year''} is an example of \textit{unconditional obligation}.

Finally, we present an example of conditional obligation where the obliged action should be performed before another event that is not a time event (there is not a deadline). \texttt{Norm07} is: ``\textit{When an agent enters into a supermarket parking between 7 a.m. and 7 p.m., they have to pay 2euro for every hour of the parking unless they did some shopping at the supermarket''}.

It is important to mention an important constraint for the use of the model: the regulated action must have an actor, this actor is the debtor of the deontic relation and is the agent who will fulfill or violate the deontic relation.


\subsection{The model of Exceptions}

The meaning of having the \textit{permission} to perform an action has been widely studied in the literature and different types of permission have been analyzed. In \cite{Governatori2013} the important distinction between \textit{strong} and \textit{weak} permission has been discussed. Having the weak permission to do an action is equivalent to the absence of the prohibition to do such an action. Differently, we have the strong permission to do an action when there is the explicit permission to do such an action; usually strong permissions are used to explicitly derogate to existing prohibitions. A similar notion is that of \textit{exemption}, which is used to derogate obligations. In the T-Norm model we introduce one construct, the \textit{exception}, that can be used for modelling both permission and exemption and can be iterated at any level of depth.

By using the T-Norm model we can specify different types of exceptions. The \textit{first type} is represented by exceptions to norms activation. When some specific conditions on the event that activates the norm are met, the consequent deontic relation has not to be generated. An example of this type of exception to \texttt{Norm01} is: \textit{``ambulances do not have to pay for entering into the limited traffic area''}. In this case a check on the type of the vehicle inhibits the creation of the obligation to pay, thus creating an exemption.

The exceptions of the \textit{second type} are those to deontic relations, i.e. when some specific conditions on the regulated event are satisfied the generation of violation/fulfilment is inhibited. An example of this type of exception to \texttt{Norm02} is: \textit{``school teachers can always borrow every DVD from the library''}. In this case the role of the borrower can be used to prevent the violation of the prohibition, that is, for creating a permission. This type of exception cannot be expressed by inhibiting the activation of the norm (using the first type of exception) because the condition (being a school teacher), is on of the borrower, who is not part of the activating event of the prohibition.

Both these types of exceptions could be expressed by adding some specific conditions to the antecedent of a norm. However, this solution requires to modify an already enforced norm by adding further conditions. This is not a good solution because it implies changing the norms while the agents may be reasoning on them. A better solution consists in expressing exceptions with a construct that is external to norms. Our idea is to introduce a construct able to inhibit the activation or the fulfilment/violation of a norm when the activating or the regulated event happens and some further conditions are met. Therefore, we formalize exceptions with a construct whose effects is to inhibit the activation of one of the components of a norm. Given that an exception is strictly related to a norm, we assume that it has access to all the variables introduced in the related norm, to which it simply adds some more conditions. An exception is expressed in one of the following ways on the basis of its type:

\footnotesize
\begin{verbatim}
EXCEPTION TO Norm_n TYPE 1
ON 	?event1
WHERE conditions on event1
THEN  exceptionToNorm(Norm_n,?event1)
\end{verbatim}
\normalsize

\footnotesize
\begin{verbatim}
EXCEPTION TO Norm_n TYPE 2
ON 	?event2
WHERE conditions on event2 AND isGenerated(?dr,Norm_n)
THEN  exceptionToDR(?dr,?event2)
\end{verbatim}
\normalsize

For example the formalization of an exception of the fist type to \texttt{Norm01} for ambulances is:
\footnotesize
\begin{verbatim}
EXCEPTION TO Norm01
ON 	 ?e1 WHERE	Ambulance(?v)
THEN exceptionToNorm(Norm01,?e1)
\end{verbatim}
\normalsize

These types of exceptions cannot be formalized by simply deleting a norm or general deontic relation, because they suspend the effects of norms only in particular situations. For example, \texttt{Norm01} applies to all vehicles except ambulances and \texttt{Norm02} applies to all library subscribers except school teachers.

By analysing real cases of norms, we realized that there exists a \textit{third type} of exceptions whose effect is to inhibit the fulfilment or violation of \textit{specific deontic relations}. These exceptions are different from those of the second type because they are triggered by an event that is not the one regulated by the norm. For example an exception to the Covid-19 \texttt{Norm03} is: \textit{``if the house is on fire then everybody is allowed to leave it''}. This exception is activated by an event (the house is on fire) that is different from the action that is regulated by the norm (leaving the house). We model those exceptions in the following way:

\footnotesize
\begin{verbatim}
EXCEPTION TO Norm_n TYPE 3
ON 	?event_n
WHERE conditions on event_n AND isGenerated(?dr, Norm_n) AND
      NOT fulfills(?agent,?dr) AND NOT violates(?agent,?dr)
THEN  exceptionToDR(?dr,?event_n)
\end{verbatim}
\normalsize

For example the exception to \texttt{Norm03} is formalized as\footnote{Where \texttt{affectedPerson} is the property that connects the event of having a positive swab with the tested person and it is used for connecting the activation event of the norm with the activation of the exception.}:
\footnotesize
\begin{verbatim}
EXCEPTION TO Norm03 TYPE 3
ON 	?en
WHERE Fire(?en) AND place(?en,?house) AND residence(?house,?agent) AND
      isGenerated(?dr,Norm03) AND activated(?dr,?e1) AND
      affectedPerson(?e1,?agent) AND NOT fulfills(?agent,?dr) AND
      NOT violates(?agent,?dr)
THEN  exceptionToDR(?dr,?en)
\end{verbatim}
\normalsize

It is also possible to have exceptions to exceptions that will inhibit the activation of the three types of exceptions described above.

\section{Operational Semantics of the model}\label{sec:semantics}
In this section, we will show how the model of norms proposed so far can be used to monitor the temporal evolution of normative states on the basis of the events occurred in the interaction among agents. Our goal is to compute the violation or fulfilment of norms on the basis of actual events.

The operational semantics of the T-NORM model can be specified by providing an unambiguous procedure for translating the model into a target formalism that already has an operational semantics. As target formalism we choose production rules, because their structure and behavior make it fairly easy to translate norms into them. Production rules, often simply called productions or rules, have been investigated in computer science, and in particular in the AI literature related to knowledge representation and reasoning \cite{10.5555/975621BrachmanLevesque2004}. A production rule has the form: \verb"IF" \textit{conditions} \verb"THEN" \textit{actions}. It has two parts: an antecedent set of \textit{conditions} that are tested on the current state of the \textit{working memory} and a \textit{consequent} set of \textit{actions} that typically modify the working memory.

The operational semantics of a production rule system is given in the W3C Recommendation of the RIF Production Rule Dialect\footnote{\url{w3.org/TR/rif-prd/#Operational_semantics_of_rules_and_rule_sets}} by means of a labeled terminal transition system. Such an operational semantics depends on the adoption of a \textit{conflict resolution strategy} for selecting the rule instance that must fire when more than one rule is applicable. Our conflict resolution strategy is as follows. Firstly, use the \textit{priority} among rules (for example, as we will discuss later, production rules for representing exceptions have higher priority than production rules for expressing norms). Secondly, when two or more rules have the same priority, use the \textit{order} conflict resolution strategy, i.e., pick the first applicable rule in order of presentation. This choice will not influence the final state reached by the working memory because the actions of the production rules used for expressing norms will never remove knowledge from the \textit{State KB}, they have effects only on the \textit{Deontic KB}. Every abstract norm of our model translates into three production rules, according to the following procedure:


\begin{enumerate}
  \item Create one production rule equal to the fist \texttt{ON}...\texttt{THEN} part in the norm. Then add the condition for managing exceptions of first type\footnote{\texttt{NOT exceptionToNorm(Norm\_n,?e1)}} and the condition for checking that the activation event happens after the creation of the norm. Then add the action for recording the instant of time when the norm is activated.
  \item Create one production rule equal to the \texttt{THEN} part of the second rule. Then add the condition for managing the exception of the second and third type\footnote{\texttt{NOT exceptionToDR(?dr,?e2) AND NOT exceptionToDR(?dr,?en)}} and the condition for checking that the regulated action is performed before \texttt{event3} (that for obligations may represent a deadline).
  \item Create one production rule for expressing the \texttt{ELSE} part of the second rule. This rule is activated when the deontic relation is not yet fulfilled and \texttt{event3} happens (for example a deadline is elapsed) and therefore the regulated action cannot be performed any more before \texttt{event3}. The procedure adds also the condition for managing the exception of the second and third type.
\end{enumerate}

The conditions of the production rules are evaluated on a \textit{working memory}, which consists of: (i) the \textit{State KB} where all the relevant events happened are recorded, those events are represented using the \textit{OWL Event Ontology}; and (ii) the \textit{Deontic KB}, where all the information for managing the evolution of the state of norms is stored. Given that the \textit{working memory} is an OWL ontology, it is possible to use OWL reasoning for computing for example that the performance of an action implies another one. This is a crucial aspect of the proposed normative model because without any further addition it is possible to reason on the effects that the logical implication between actions has on norms fulfilment or violation. In fact, we obtain that: (i) the obligation to perform an action is fulfilled by any action that implies the regulated one. This is because we have the following chain of implications: action a1 implies action a2 and a2 produces a fulfilment, so a1 leads to a fulfilment. (ii) Similarly, the prohibition to perform an action is violated by the performance of any action that implies the regulated one. (iii) The permission to perform a generic action implies the permission to perform all the more specific actions implied by the generic one. This is because the specific action implies the more generic one that will activate the exception that in turn inhibits the norm. For example if an agent has the permission to transfer the property of a product, thanks to OWL reasoning he has also the permission to sell or to give to someone else the product.


In the proposed framework for norms monitoring, we take advantage of two types of computation: OWL reasoning on the state of the interaction and forward chaining realized by means of production rules. OWL reasoning and forward chaining are combined in a safe manner because they alternate. In fact, the main loop of a software, able to monitor or simulate the evolution of the norms state over time, is as follows:
\begin{enumerate}
  \item Every time new events are added to the \textit{State KB}, execute an OWL reasoner. We assume that only events happening in the future (for simulation) or at the current time (for monitoring) may be inserted in \textit{State KB};
  \item Store the resulting \textit{State KB} together with the \textit{Deontic KB} in the working memory of a production system and execute the production rule engine;
  \item Wait for the next event, in monitoring, or move the simulation time to next significant instant of time and go back to point 1).
\end{enumerate}

In our model we need to combine those two types of computation because it is not possible to use only OWL reasoning for computing the violation or fulfilment of norms. In fact, when norms regulate time constrained actions, it is necessary to deduce that the non-performance of the regulated action before a deadline implies violation or a fulfilment. Given that OWL reasoning works on an open world assumption, inferences of this type cannot be drawn directly. One possible solution to this problem is computing the closure of specific classes using an external routine like it is proposed in \cite{DBLP:journals/aicom/FornaraColombetti10}. The advantage of using production rules is a clear separation of two different types of computation, each one used coherently with its nature, and having a more declarative solution where the semantics is expressed with production rules instead of using Java code.

Every exception translates into one production rule. Given that an exception has access to all the variables introduced in the related norm the conditions in the norm are automatically merged with the condition of the exception. First and second type exceptions must fire before norms therefore their productions have a higher \textit{priority} than norms and exceptions to exceptions have higher priority than exceptions to norms. Thanks to the fact that exception production rules will act before the norm production rules, the exception will settle false one of the conditions of the norm that therefore will not act any more.

\vspace{-0,3cm}
\subsection{Implementation of a Prototype}\label{sec:prototype}

We tested the described framework with a Java prototype that uses Pellet\footnote{\url{http://pellet.owldl.com/}}, an open-source Java based OWL 2 reasoner, and the JENA general purpose rule engine\footnote{\url{https://jena.apache.org/documentation/inference/\#rules}} for realizing forward chaining on production rules. The reason why we chose to use the JENA framework is that, differently from other rule-based systems like DROOLS (used in \cite{DBLP:conf/atal/Alvarez-NapagaoAVD10}) or Jess (used in \cite{Garcia-Camino:2005:INE:1082473.1082575}), its rule engine natively supports rule-based computations over an OWL ontology serialized as an RDF graph. JENA provides forward chaining realized by means of an internal RETE-based interpreter. To test the framework, the various type of norms discussed in Section \ref{sec:expressive:power} were translated manually into the syntax of JENA production rules.

\vspace{-0,3cm}
\section{Related Work}\label{sec:related:work}

In the literature, there are various proposals where a model of norms and policies is formalized using different languages and where different frameworks are investigated with the goal of providing various services. Useful services are: searching of policies having certain characteristics \cite{DBLP:journals/semweb/OltramariPSWCNR18}, anticipating conflicts among policies \cite{DBLP:journals/ws/SensoyNVS12}, monitoring \cite{DBLP:journals/aicom/FornaraC19} or compliance checking \cite{2019:ODRL:DeVosKirranePadget}, and simulation for performing a what-if reasoning \cite{KAOS:2008}.


One of the pioneer techniques for normative reasoning is deontic logic \cite{vonWright1951}. Despite deontic logic approaches present some limitations, for example the triggering and regulated actions are usually expressed with simple propositional formulae \cite{Governatori:2010:10.5555/1862330.1862332}, some of their basic concepts and insights are still used in many recent approaches where other formal languages are used. In order to pursue interoperability among different normative systems, it is crucial to use a standard language for the formalization of norms. Today's there are two standards: the already mentioned ODRL policy expression language which is a W3C Recommendation and the OASIS standard LegalRuleML\footnote{\url{https://www.oasis-open.org/committees/legalruleml/}} that defines a rule interchange language for the legal domain and is formalized using RuleML. ODRL has many connections with the model proposed in this paper as it specify a language for expressing obligations, prohibitions, and permissions. A great limitation of ODRL is not having an operational semantics that allows to compute the fulfillment or the violation of policies. In our previous work \cite{DBLP:journals/aicom/FornaraC19} we proposed to extend the ODRL information model for being able to express its operational semantics using finite state machines implemented using production rules. In this work we have moved further away from ODRL, because we would like to overcome some limitations. Firstly, in ODRL it is not possible to specify generic policies applicable every time to a different agent. In ODRL the debtor of a policy can only be a specific agent. Differently, thanks to our abstract model for policies specification, it is possible to apply one policy to all the agents who will perform a certain action (for example having a positive swab) or who plays a certain role. Secondly, we don't consider exceptions (and in particular permissions) at the same level of obligations and prohibitions. From our prospective exceptions are derived concepts and they exist only if there is a corresponding basic level construct that expresses obligations and prohibitions. Finally, ODRL has a fixed set of deontic types, in our model our focus is on specifying the sequence of events that bring to a violation or to a fulfillment. An important aspect that the T-NORM and the ODRL model have in common is the use of semantic web technologies. Although they use them for different purposes: ODRL uses the OWL language for the specification of the policy meta-model, while in the T-NORM model, as well as in the in the OWL-POLAR model \cite{DBLP:journals/ws/SensoyNVS12}, the OWL language is used for modeling the actions performed by agents and consequently to express the activation conditions and the actions regulated by the norms.

We now continue our comparison by focusing on models of norms that are expressed using semantic web technologies and productions, even if none of them combine the two technologies as we do. Two features that make our model innovative are: the formalization of the relation between norms and time constraints and the possibility to directly describe which sequence of events or actions would bring an agent to the violation or to the fulfilment of a norm. In the OWL-POLAR framework \cite{DBLP:journals/ws/SensoyNVS12}, similarly to our approach, the state of the world is represented using an OWL ontology. Differently, policies activation is computed by translating the conjunctive semantic formulae, used for describing what is prohibited, permitted or obliged by the policy, into SPARQL queries that are evaluated on the state of the world. In this work we propose a more straight approach where norms conditions are directly evaluated on the state of the world without the need of translations. An interesting aspect of the OWL-POLAR framework that we plan to investigate in our future works, is the mechanism for anticipating possible conflicts among policies, and for conflicts avoidance and resolution.

In \cite{DBLP:conf/atal/Alvarez-NapagaoAVD10} one type of norm is defined as a tuple that can generate a norm instance that in turn can be fulfilled or violated. A norm specifies a target condition that describes the state that fulfill the norm and a maintenance condition used for defining the conditions that, when no longer hold, lead to a violation. In this approach count-as rules are used to introduce institutional facts, regulated by norms, starting from brut events. Differently from our model, where it is possible to model different type of deadlines, in this approach only a timeout property, i.e. a deadline for the reparation of the violation of a norm, is taken into account. In \cite{DBLP:conf/atal/Alvarez-NapagaoAVD10} an interesting violation handling norm is formalized that is activated when another norm is violated. Similarly to our approach the monitoring is realized using a production system that concretely is realized using DROOLS, but no discussion is tackled on the advantages of using OWL reasoning and on how to combine it with forward-chaining realized by means of production rules.

Another interesting proposal is the KAoS policy management framework \cite{DBLP:conf/policy/UszokBLBBFJJ08}. In KAoS Semantic Web
technologies are used for policy specification and management, in particular policy monitoring and enforcing is realized by a component that compiles OWL policies into an efficient format. In literature there are other interesting approaches where norms are specified as rules but they are not taking advantage of the use of semantic web technologies. For example in \cite{10.1007/11775331_10:Vigano:2006}  norms are generators of commitments for the agents playing a certain role in an artificial institution. In \cite{Garcia-Camino:2005:INE:1082473.1082575} norms have a type and may have a deadline. Thanks to their form of preconditions $\rightarrow$ postconditions are easily expressible with Jess rules\footnote{Jess is a rule engine for the Java platform}. Finally in \cite{2019:ODRL:DeVosKirranePadget} an extension of the ODRL language is proposed to capture the semantics of business policies thanks to their translation into Answer Set Programming for making it possible to realize compliance checking. An interesting aspect of this work is that the result of compliance checking can be positive or negative with an explanation of the aspects of the policy that caused the non-compliance.


In our future work, we plan to investigate the application of sanctions or rewards and to study the formalization of the notion of institutional power. We plan to further investigate the expressive power of the model for specifying other deontic relations.

\vspace{0,3cm}
\footnotesize
\noindent \textbf{Acknowledgement}

The research reported in this paper has been funded by the SNSF (Swiss National Science Foundation) grant no. 200021\_175759/1.
We acknowledge the contribution to this research by Mr Marco Sterpetti during his master thesis at Politecnico
di Milano.
\normalsize

\vspace{-0,3cm}
\bibliographystyle{abbrv}
\bibliography{TimeRelatedNorms}

\begin{thebibliography}{10}

\bibitem{DBLP:conf/atal/Alvarez-NapagaoAVD10}
S.~{\'{A}}lvarez{-}Napagao, H.~Aldewereld, J.~V{\'{a}}zquez{-}Salceda, and
  F.~Dignum.
\newblock Normative monitoring: Semantics and implementation.
\newblock In M.~D. Vos et~al., editors, {\em COIN@AAMAS 2010, Toronto, Canada,
  May 2010, COIN@MALLOW 2010, Lyon, France, August 2010, Revised Selected
  Papers}, volume 6541 of {\em LNCS}, pages 321--336. Springer, 2010.

\bibitem{10.5555/975621BrachmanLevesque2004}
R.~Brachman and H.~Levesque.
\newblock {\em Knowledge Representation and Reasoning}.
\newblock Morgan Kaufmann Publishers Inc., San Francisco, CA, USA, 2004.

\bibitem{2019:ODRL:DeVosKirranePadget}
M.~De~Vos, S.~Kirrane, J.~Padget, and K.~Satoh.
\newblock Odrl policy modelling and compliance checking.
\newblock In P.~Fodor, M.~Montali, D.~Calvanese, and D.~Roman, editors, {\em
  Rules and Reasoning}, pages 36--51, Cham, 2019. Springer International
  Publishing.

\bibitem{fornara:SASFA:2011}
N.~Fornara.
\newblock {Specifying and Monitoring Obligations in Open Multiagent Systems
  using Semantic Web Technology}.
\newblock In {\em {Semantic Agent Systems: Foundations and Applications}},
  volume 344 of {\em Studies in Computational Intelligence}, chapter~2, pages
  25--46. Springer-Verlag, 2011.

\bibitem{DBLP:journals/aicom/FornaraColombetti10}
N.~Fornara and M.~Colombetti.
\newblock {Representation and monitoring of commitments and norms using OWL}.
\newblock {\em AI Commun.}, 23(4):341--356, 2010.

\bibitem{DBLP:journals/aicom/FornaraC19}
N.~Fornara and M.~Colombetti.
\newblock Using semantic web technologies and production rules for reasoning on
  obligations, permissions, and prohibitions.
\newblock {\em {AI} Commun.}, 32(4):319--334, 2019.

\bibitem{Garcia-Camino:2005:INE:1082473.1082575}
A.~Garcia-Camino, P.~Noriega, and J.~A. Rodriguez-Aguilar.
\newblock Implementing norms in electronic institutions.
\newblock In {\em Proceedings of the Fourth International Joint Conference on
  Autonomous Agents and Multiagent Systems}, AAMAS '05, pages 667--673, New
  York, NY, USA, 2005. ACM.

\bibitem{Governatori2013}
G.~Governatori, F.~Olivieri, A.~Rotolo, and S.~Scannapieco.
\newblock Computing strong and weak permissions in defeasible logic.
\newblock {\em Journal of Philosophical Logic}, 42(6):799--829, Dec 2013.

\bibitem{Governatori:2010:10.5555/1862330.1862332}
G.~Governatori and A.~Rotolo.
\newblock A conceptually rich model of business process compliance.
\newblock In {\em Proceedings of the Seventh Asia-Pacific Conference on
  Conceptual Modelling - Vol. 110}, pages 3–--12. Australian Computer
  Society, Inc., 2010.

\bibitem{10.1007/11564751_2Horrocks:OWL}
I.~Horrocks.
\newblock {OWL: A Description Logic Based Ontology Language}.
\newblock In P.~van Beek, editor, {\em Principles and Practice of Constraint
  Programming - CP 2005}, pages 5--8, Berlin, Heidelberg, 2005. Springer Berlin
  Heidelberg.

\bibitem{DBLP:journals/semweb/OltramariPSWCNR18}
A.~Oltramari, D.~Piraviperumal, F.~Schaub, S.~Wilson, S.~Cherivirala, T.~B.
  Norton, N.~C. Russell, P.~Story, J.~R. Reidenberg, and N.~M. Sadeh.
\newblock Privonto: {A} semantic framework for the analysis of privacy
  policies.
\newblock {\em Semantic Web}, 9(2):185--203, 2018.

\bibitem{Panagiotidi:2013:TNA:2731087.2731112}
S.~Panagiotidi, S.~Alvarez-Napagao, and J.~V\'{a}zquez-Salceda.
\newblock Towards the norm-aware agent: Bridging the gap between deontic
  specifications and practical mechanisms for norm monitoring and norm-aware
  planning.
\newblock In {\em Revised Selected Papers of the COIN 2013}, volume 8386, pages
  346--363. Springer-Verlag Inc., 2014.

\bibitem{DBLP:journals/ws/SensoyNVS12}
M.~Sensoy, T.~J. Norman, W.~W. Vasconcelos, and K.~P. Sycara.
\newblock {OWL-POLAR:} {A} framework for semantic policy representation and
  reasoning.
\newblock {\em J. Web Sem.}, 12:148--160, 2012.

\bibitem{KAOS:2008}
A.~{Uszok}, J.~M. {Bradshaw}, J.~{Lott}, M.~{Breedy}, L.~{Bunch},
  P.~{Feltovich}, M.~{Johnson}, and H.~{Jung}.
\newblock {New Developments in Ontology-Based Policy Management: Increasing the
  Practicality and Comprehensiveness of KAoS}.
\newblock In {\em 2008 IEEE Workshop on Policies for Distributed Systems and
  Networks}, pages 145--152, 2008.

\bibitem{DBLP:conf/policy/UszokBLBBFJJ08}
A.~Uszok, J.~M. Bradshaw, J.~Lott, M.~R. Breedy, L.~Bunch, P.~J. Feltovich,
  M.~Johnson, and H.~Jung.
\newblock {New Developments in Ontology-Based Policy Management: Increasing the
  Practicality and Comprehensiveness of KAoS}.
\newblock In {\em {POLICY} 2008, 2-4 June 2008, Palisades, New York, {USA}},
  pages 145--152. {IEEE} Computer Society, 2008.

\bibitem{10.1007/11775331_10:Vigano:2006}
F.~Vigan{\`o}, N.~Fornara, and M.~Colombetti.
\newblock An event driven approach to norms in artificial institutions.
\newblock In O.~Boissier et~al., editors, {\em Coordination, Organizations,
  Institutions, and Norms in Multi-Agent Systems. AAMAS 2005}, volume 3913 of
  {\em LNCS}, pages 142--154. Springer Berlin Heidelberg, 2006.

\bibitem{NORMAS:2018}
G.~E.~S. Villata, editor.
\newblock {\em Special Issue: Normative Multi-Agent Systems}, volume~5 of {\em
  Journal of Applied Logics - IfCoLog Journal}. College Publications, 2018.

\bibitem{vonWright1951}
G.~H. von Wright.
\newblock Deontic logic.
\newblock {\em Mind, New Series}, 60(237):1--15, 1951.

\end{thebibliography}

\end{document}